\documentclass[12pt]{article}

\usepackage{graphicx}
\usepackage{a4}

\usepackage{fancybox}
\usepackage{epsfig}
\usepackage{color}

\usepackage{amsfonts}
\usepackage{mathrsfs}
\usepackage{amssymb}
\usepackage{amsmath}

\usepackage{dcolumn}
\usepackage{bm}

\def\pa{\partial}

\def\al{\alpha}

\newcommand{\ben}{\begin{equation}}
\newcommand{\een}{\end{equation}}
\newcommand{\bea}{\begin{eqnarray}}
\newcommand{\eea}{\end{eqnarray}}
\newcommand{\ba}{\begin{array}}
\newcommand{\ea}{\end{array}}
\newcommand{\bit}{\begin{itemize}}
\newcommand{\eit}{\end{itemize}}

\newcommand{\vs}[1]{\vspace{#1 mm}}

\newcommand{\dsl}{\pa \kern-0.5em /}

\begin{document}

\topmargin 0pt \oddsidemargin 0mm

\begin{flushright}

USTC-ICTS-10-10\\

\end{flushright}

\vspace{2mm}

\begin{center}

{\Large \bf Cosmological evolution of a D-brane}

\vs{10}

 {\large Huiquan Li \footnote{E-mail: hqli@ustc.edu.cn}}

\vspace{6mm}

{\em

Interdisciplinary Center for Theoretical Study\\

University of Science and Technology of China, Hefei, Anhui 230026, China\\

}

\end{center}

\vs{9}

\begin{abstract}
We study the cosmological evolution of a single BPS D-brane coupled
to gravity in the absence of potential. When such a D-brane moves in
the bulk with non-vanishing velocity, it tends to slow down to zero
velocity via mechanisms like gravitational wave leakage to the bulk,
losing its kinetic energy to fuel the expansion of the universe on
the D-brane. If the initial velocity of the D-brane is high enough,
the universe on the D-brane undergoes a dust-like stage at early
times and an acceleration stage at late times, realising the
original Chaplygin gas model. When the D-brane velocity is initially
zero, the D-brane will always remain fixed at some position in the
bulk, with the brane tension over the Plank mass squared as a
cosmological constant. It is further shown that this kind of fixed
brane universe can arise as defects from tachyon inflation on a
non-BPS D-brane with one dimension higher.
\end{abstract}


\section{Introduction}
\label{sec:introduction}

In past years, various dark energy models have been proposed to
explain the accelerated expansion of the universe. Among them, the
simplest way is to introduce a non-vanishing cosmological constant.
However, this model encounters a fine tuning problem in explaining
the smallness of the cosmological constant. Other than introducing a
constant, there are dynamical dark energy models, like the Chaplygin
gas model, which potentially provides a unified description of dark
matter and dark energy \cite{Kamenshchik:2001cp,Bilic:2001cg}. In
this model, the acceleration of the universe is explained to be due
to the cosmological evolution of some Chaplygin gas whose energy
density and the pressure satisfy: $\rho p=-A$ ($A>0$). This energy
density and pressure relationship can be interpreted as a D-brane
moving in the bulk in the absence of potential, whose dynamics is
described by a Dirac-Born-Infeld (DBI) effective action. In the
homogeneous and isotropic Friedmann-Robertson-Walker (FRW)
cosmology, the general continuity equation is $d(\rho
a^3)=-pd(a^3)$, with $a$ the scale factor. From this equation, the
evolutionary rule of the energy density of the Chaplygin gas was
derived:
\begin{equation}\label{e:Chapgasevol}
 \rho=\sqrt{A+\frac{B}{a^6}},
\end{equation}
where $B$ is an integral constant. Hence, the energy density
$\rho\propto a^{-3}$ at early times and $\rho\propto\sqrt{A}$ at
late times. The two stages therefore are suggested to account for
dark matter and dark energy in cosmology, respectively. However,
this model gets negative support from observations
\cite{Sandvik:2002jz,Amendola:2003bz} because the perturbation
spectrum tends to vanish at the acceleration stage.

In this paper, we examine in detail the dynamics of a single BPS
D-brane coupled to gravity in the absence of potential. We show that
the Chaplygin gas can indeed be interpreted as such a D-brane
slowing down from a high speed in the bulk, with the kinetic energy
transferred to the internal energy of the brane to fuel the
expansion of the universe on it. However, in using this
interpretation, there is a problem that needs to be solved first:
what is the mechanism that initially triggers the slow down of the
D-brane? In this paper, we argue that a mechanism may be the
gravitational wave leakage from the brane universe to the bulk.

The investigation on the evolution of a moving D-brane in the
absence of potential is useful for understanding the dynamics of
D-branes that have attained velocities via different mechanisms in
various brane-based cosmological models. In the mirage cosmology
model \cite{Kehagias:1999vr}, the brane attains velocity through
interactions with various fields in the bulk or on other branes. In
the brane inflation model
\cite{Burgess:2001fx,Jones:2002cv,Kachru:2003sx}, a pair or a stack
of D-branes and anti-D-branes can roll slowly down the potential,
giving rise to inflation, while, in the DBI inflation model
\cite{Silverstein:2003hf,Alishahiha:2004eh}, the D-branes can roll
fast in the bulk. In the tachyon inflation model
\cite{Gibbons:2002md,Fairbairn:2002yp,Choudhury:2002xu,Frolov:2002rr,
Kofman:2002rh,Steer:2003yu}, a single non-BPS D-brane can also
attain velocity driven by the asymptotic tachyon potential. Our
results obtained in this paper indicate that the moving D-branes in
these models will eventually slow down to zero velocity if the
potentials become unimportant at the end of inflation.

Besides the above case for a moving brane, we also find that the
D-brane can be kept fixed at some position in the bulk if its
initial velocity is zero. This case is similar to the DGP model
\cite{Dvali:2000hr,Deffayet:2000uy,Deffayet:2001pu}, in which a
three-brane universe is assumed to be embedded in a five-dimensional
Minkowski bulk at a fixed position. Here, we verify that a brane
indeed can remain fixed in the bulk as the universe on it expands.
Moreover, this kind of fixed D-brane can originate from tachyon
inflation on a non-BPS D-brane with one dimension higher.

This paper is organised as follows. In Sec. 2, we present the set-up
of the model. In Sec. 3 and 4, we discuss the two cosmological
evolution solutions depending on whether the initial velocities of
the D-brane is non-zero or zero. The last section gives the
conclusion.

\section{The universe on a D-brane}
\label{sec:braneuni}

D3-branes in Type IIB superstring theory are BPS supersymmetry
saturated and stable. In this paper, we consider the dynamics of a
single BPS D3-brane in the 10-dimensional bulk, with six extra
dimension compactified, in the absence of potential. Concretely, we
consider the D3-brane to move in a FRW metric in the four
uncompactified bulk along which the brane extends. Note that the
tension of a BPS D3-brane is: $\tau_3=1/[(2\pi)^3g_sl_s^4]$, where
$g_s$ is the string coupling and $l_s=\sqrt{\al'}$ is the string
length. The Plank mass on the four uncompactified bulk can be
obtained from the 10-dimensional one via dimensional reduction
\cite{Jones:2002cv,Kofman:2002rh}: $M_{Pl}^2=V/(g_sl_s)^2$, where
$V=r_0^6/(\pi l_s^6)$ is the volume of the 6-dimensional
compactified space and $r_0$ is the radius of the compactified
space. Based on them, we introduce the following parameter which is
useful for later discussions:
\begin{equation}\label{e:lambdasqr}
 \lambda^2=\frac{\tau_3}{3M_{Pl}^2}=\frac{g_s}{3(2\pi)^2l_s^2V}.
\end{equation}
Its magnitude depends only on $g_s$ for fixed $\al'$ and $V$.
According to \cite{Lu:2007kv}, a D3-brane affects the bulk strongly
only within a distance from the brane: $r\lesssim g_s^{1/4}l_s$,
which is even shorter than the string length $l_s$ for a small $g_s$
or $\lambda$. In our following discussion, $\lambda$ is indeed
expected to be small.

On a D$3$-brane, we have the low-energy fields including the metric
$g_{\mu\nu}$ ($\mu,\nu=0,1,2,3$), the $U(1)$ gauge fields and the
massless scalars $Y^I$ ($I=4,5,\cdots,9$) that describe the motion
of the D3-brane in the six transverse spatial directions. For
simplicity, we take into account no gauge fields. Denoting the
transverse scalars as a vector $\vec{Y}=\{Y^I\}$, we can write
$\partial_\mu\vec{Y}\cdot\partial_\nu\vec{Y}=\sum_I
\partial_\mu Y^I\partial_\nu Y^I$. Since the direction of
$\vec{Y}$ is not important in our discussion, we only consider the
value $\partial_\mu Y\partial_\nu Y$ of
$\partial_\mu\vec{Y}\cdot\partial_\nu\vec{Y}$. Under this
consideration, the DBI style action for a D3-brane coupled to
gravity is given by
\begin{eqnarray}\label{e:CosMTDBIA}
 S=\frac{1}{16\pi G}\int d^4x\sqrt{-g}R-\tau_3\int d^4x\sqrt{-\det
\left(g_{\mu\nu}+\partial_\mu Y\partial_\nu Y\right)}
\nonumber \\
=\frac{1}{16\pi G}\int d^4x\sqrt{-g}\left[R-16\pi G\tau_3\sqrt{
1+g^{\mu\nu}\partial_\mu Y\partial_\nu Y}\right].
\end{eqnarray}

We assume that it is possible to obtain an induced FRW metric on the
D3-brane: $ds^2=-dt^2+a^2(t)d\vec{x}^2$ and as usual $Y$ to depend
on time only, being homogeneous and isotropic. The Einstein
equations lead to the following Friedmann equation
\begin{equation}\label{e:friedman1}
 H^2=\left(\frac{\dot{a}}{a}\right)^2=\frac{8\pi G}{3}
\rho-\frac{K}{a^2}.
\end{equation}
The energy density and the pressure on the brane are respectively:
\begin{equation}\label{e:enpre}
 \rho=\frac{\tau_3}{\sqrt{1-\dot{Y}^2}}, \textrm{ }
\textrm{ }\textrm{ } p=-\tau_3\sqrt{1-\dot{Y}^2}.
\end{equation}
Thus, we have negative pressure on the D-brane for $|\dot{Y}|<1$ and
vanishing pressure for $|\dot{Y}|=1$. The pressure and the energy
density satisfy $\rho p=-\tau_3^2$, for which D-branes moving in the
bulk can be viewed as a kind of Chaplygin gas
\cite{Kamenshchik:2001cp,Bilic:2001cg}. The equation of state is
$w=p/\rho=-(1-\dot{Y}^2)\in[-1,0]$.

In the flat universe ($K=0$), Eq.\ (\ref{e:friedman1}) is
\begin{equation}\label{e:friedman11}
 H^2=\frac{\lambda^2}{\sqrt{1-\dot{Y}^2}},
\end{equation}
where $\lambda^2=\tau_3/(3M_{Pl}^2)$, as given in Eq.\
(\ref{e:lambdasqr}), and the Plank mass is $M_{Pl}=1/\sqrt{8\pi G}$
here. We can also have another constraining equation from the
equation of motion of $Y$ in the FRW metric
\begin{equation}\label{e:eomY}
 \frac{\ddot{Y}}{1-\dot{Y}^2}+3H\dot{Y}=0.
\end{equation}
In the next two sections, we will discuss two kinds of solutions to
the above two equations, depending on whether the D-brane is
initially moving or fixed in the bulk.

\section{Moving brane: the Chaplygin gas model}
\label{sec:chaplygingas}

For an expanding universe with positive $H$, the first solutions
derived from Eqs.\ (\ref{e:friedman11}) and (\ref{e:eomY}) are
\begin{equation}\label{e:mastereq}
 \ln\sqrt{\frac{H+\lambda}{H-\lambda}}+\arctan\left(
\frac{\lambda}{H}\right)=3\lambda (t+t_0),
\end{equation}
\begin{equation}\label{e:Hsol}
 H=\frac{\lambda}{(1-\dot{Y}^2)^{\frac{1}{4}}},
\end{equation}
where $t_0$ is a constant. 
It is clear that $H$ and $\dot{Y}^2$ in the solutions are both
monotonic functions of $t$, decreasing with the time growing.
Specially, when $t_0=0$, $\dot{Y}^2=1$, $H=\infty$ at $t=0$ and
$\dot{Y}^2=0$, $H=\lambda$ at $t=\infty$. Therefore, the speed
$|\dot{Y}|$ of the D3-brane decreases while the scale factor $a(t)$
increases as time grows, behaving as the expanding universe.
The universe on the D3-brane evolves from a pressureless dust-like
state with $\omega=0$ to a vacuum with $\omega=-1$ as time grows.

Without loss of generality, we set the time to run from $t=0$ to
$t=\infty$. So the constant $t_0$ decides the initial values of
$\dot{Y}(t=0)$ and $H(t=0)$. In the special case of $t_0=0$, the
initial speed $|\dot{Y}|$ is equal to the critical value $1$ and so
the energy density $\rho$ and the Hubble parameter $H$ are both
singular at the beginning of time.

\subsection{Dust-like era and late-time acceleration}

Let us take a look at the above solutions at different epochs. At
early times, $H$ is large $H\gg\lambda$ when $t_0\ll1$. In this
case, the left side of Eq.\ (\ref{e:mastereq}) approximately equals
$2\lambda/H$ and the solution can be written
\begin{equation}\label{e:earlyappHsol}
 H\simeq\frac{2}{3(t+t_0)}.
\end{equation}
Thus, the scale factor and the energy density evolves as,
\begin{equation}
 a(t)
\simeq\left(1+\frac{t}{t_0}\right)^{\frac{2}{3}}\propto
(t+t_0)^{2/3}, \textrm{ }\textrm{ }\textrm{ } \rho(t) \propto a^{-3}
\end{equation}
This is the evolutionary rule for dust in the expanding universe.
This is understandable because the equation of state $\omega\sim0$
as $|\dot{Y}|\sim1$ at early times. This dust-like phase shares a
common sense with the pressureless tachyon matter, which forms as
the time derivative of the tachyon $|\dot{T}|\rightarrow1$ at the
end of tachyon inflation on a non-BPS D3-brane. The tachyon matter
was investigated as a candidate to cold dark matter
\cite{Sen:2002in,Sen:2002an,Frolov:2002rr}.

At late times, the second term on the left-hand side of Eq.\
(\ref{e:mastereq}) is negligible since it can only vary within
$[0,\pi/4]$. Thus, the solution of $H$ reduces to
\begin{equation}\label{e:laterappHsol}
 H\simeq\lambda\coth[3\lambda (t+t_0)],
\end{equation}
So $H$ decreases to a constant $H\simeq\lambda$ at late times,
yielding an exponentially expanding universe $a(t)\propto e^{\lambda
t}$ with $\rho\propto\lambda^2$.

Between these two stages, there should be a critical time, at which
the universe turns from a deceleration stage to an acceleration
stage. Let us now determine this critical time. From Eqs.\
(\ref{e:friedman11}) and (\ref{e:eomY}), we can derive
\begin{equation}\label{e:friedman2}
 \frac{\ddot{a}}{a}
=\lambda^2\frac{1-\frac{3}{2}\dot{Y}^2}{\sqrt{1-\dot{Y}^2}}.
\end{equation}
It is easy to know that when $\dot{Y}^2$ decreases to
$\dot{Y}^2=2/3$, i.e., the time reaches
\begin{equation}\label{e:t_c}
 t_c=\frac{0.55}{\lambda}-t'_0,
\end{equation}
the universe enters an acceleration stage from a deceleration stage.
The equation of state is $\omega>-1/3$ at the deceleration stage
($t<t_c$) and $\omega<-1/3$ at the acceleration stage ($t>t_c$).

The above results are consistent with those observed from Eq.\
(\ref{e:Chapgasevol}) in the original Chaplygin gas model
\cite{Kamenshchik:2001cp,Bilic:2001cg}, which is proposed to give a
unified description of dark matter and dark energy. However, this
model has got negative support from observations
\cite{Sandvik:2002jz,Amendola:2003bz}.

\subsection{The trigger of deceleration in the bulk}

In the above analysis as well as in the original Chaplygin gas model
\cite{Kamenshchik:2001cp,Bilic:2001cg}, a problem is ignored. That
is, the reason that the D-brane automatically decelerates in the
bulk is unclear. If we consider a system including a potential in
the bulk, the slow down of the whole D-brane may be explained as a
result of the brane-bulk or brane-brane interactions. In the system
concerned here, we may think that a possible mechanism could be the
gravitational wave leakage from the brane universe to the bulk. As
analyzed in \cite{Fabris:2004dp}, there indeed are gravitational
waves produced in the Chaplygin gas model and the evolution of the
gravitational waves does not obviously distinguish from some other
models, like the $\Lambda$CDM model. Since gravity can propagate in
the 10-dimensional bulk, the gravitational waves produced on the
D-brane are possible to flee off the brane, taking away part of the
kinetic energy and triggering the deceleration of the D-brane. Of
course, most of the kinetic energy lost due to the deceleration is
transferred to fuel the expansion of the brane universe such that
Eqs.\ (\ref{e:friedman11}) and (\ref{e:eomY}) are still correct.

The gravitational waves are associated with the tensor perturbations
$h_{ij}$, which are symmetric and traceless. Following
\cite{Liddle:1993fq,Mukhanov:2005sc}, the gravitational waves evolve
as in the universe with perfect fluid
\begin{equation}\label{e:gweqn}
 \ddot{h}_{ij}+3H\dot{h}_{ij}-\frac{1}{a^2}\vec{\nabla}^2h_{ij}=0.
\end{equation}
Defining $h_{ij}=a^{-3/2}\sum_{\vec{k}}h_{\vec{k}}\exp(i\vec{k}\cdot
\vec{x})e_{ij}$, we can rewrite the perturbation equation as
\begin{equation}\label{e:regweqn}
 \ddot{h}_{\vec{k}}+\left(\frac{\vec{k}^2}{a^2}-f(\dot{Y})
\right)h_{\vec{k}}=0.
\end{equation}
where
\begin{equation}\label{e:fdoty}
 f(\dot{Y})=\frac{3}{2}\frac{\ddot{a}}{a}+\frac{3}{4}H^2
=\frac{9\lambda^2}{4}\sqrt{1-\dot{Y}^2}.
\end{equation}
To get propagating gravitational waves, one needs
$\vec{k}^2/a^2>f(\dot{Y})$. At early times as $|\dot{Y}|\sim1$,
there exist gravitational waves for almost all wavelengths. With
time growing, the scale factor increases and the velocity
$|\dot{Y}|$ decreases. Hence, the metric perturbations with shorter
wavelengths tend to freeze out and there are now less gravitational
waves to leak to the bulk when the velocity of the D-brane
decreases. This basically agrees with the conjecture given above
that gravitational leakage triggers the slow down of the D-brane.

\section{Fixed brane universe}
\label{sec:}

There exist another kind of solutions to Eqs.\ (\ref{e:friedman11})
and (\ref{e:eomY}). In terms of Eq.\ (\ref{e:friedman11}), we can
write Eq.\ (\ref{e:eomY}) as
\begin{equation}\label{e:accofy}
 \ddot{Y}=-3\lambda\dot{Y}(1-\dot{Y}^2)^{\frac{3}{4}}.
\end{equation}
It is clear that, in the special cases $|\dot{Y}|=0$ and $=1$,
$\ddot{Y}=0$. This means that the speed of the D-brane does not
change if its velocity is zero or the critical value $1$. That is,
the brane will remain fixed at some position in the bulk if its
initial velocity is zero, while it will keep moving with the same
velocity $\dot{Y}=\pm1$ if its initial velocity is $\dot{Y}=\pm1$.
In these cases, the resulting Hubble parameter $H$ is constant.

As addressed earlier, the energy density $\rho$ and the Hubble
parameter $H$ on the D-brane with non-vanishing tension $\tau_3$ are
singular as the D-brane moves at the critical speed $\pm1$.
Therefore, we only consider the case of $\dot{Y}=0$, in which the
action (\ref{e:CosMTDBIA}) reduces to
\begin{equation}\label{e:fixbranuni}
 S=\frac{1}{16\pi G}\int d^4x\sqrt{-g}(R-6\lambda^2).
\end{equation}
It describes a universe with a positive cosmological constant, in
the absence of other matter fields. This is a special evolutionary
mode that was not included in the original Chaplygin gas model.

\subsection{Evolution of perturbations}

Although the brane is fixed at some position in the bulk, it can
still fluctuate. In what follows, we discuss the fluctuations of
such a brane universe. We assume that there are small perturbations
of the form $Y=Y(t)+y(t,x)$ in the homogeneous and isotropic brane
universe background. When the brane fluctuates around the fixed
position of the brane, the action (\ref{e:CosMTDBIA}) is
\begin{equation}\label{e:perfixbraneuni}
 S=\frac{1}{16\pi G}\int d^4x\sqrt{-g}\left[R-6\lambda^2
\sqrt{1+g^{\mu\nu}\partial_\mu y\partial_\nu y}\right].
\end{equation}
The equation of state derived from this action is
\begin{equation}\label{e:eosperfix}
 \omega=\frac{p}{\rho}=-\left[1-\frac{a^2\dot{y}^2
+\frac{1}{3}(\vec{\nabla}y)^2}{a^2+(\vec{\nabla}y)^2}\right].
\end{equation}

Correspondingly, there are the scalar metric perturbations in the
FRW cosmological background:
$ds^2=(1+2\Phi)dt^2-(1-2\Phi)a^2(t)d\vec{x}^2$. The equations for
the perturbations $y$ and $\Phi$ are respectively
\begin{equation}\label{e:pereqny}
 \ddot{y}+3H\dot{y}-\vec{\nabla}^2y=0,
\end{equation}
\begin{equation}\label{e:pereqnphi}
 \dot{\Phi}+H\Phi=0.
\end{equation}
Thus, $\Phi\propto a^{-1}$ drops to zero as the universe expands.
Similarly, we define
$y=a^{-3/2}\sum_{\vec{q}}y_{\vec{q}}\exp(i\vec{q}\cdot\vec{x})$. The
first equation above can be rewritten as
\begin{equation}\label{e:repereqny}
 \ddot{y}_{\vec{q}}+\left[\vec{q}^2-\left(\frac{3}{2}\frac{\ddot{a}}
{a}+\frac{3}{4}H^2\right)\right]y_{\vec{q}}=0.
\end{equation}

In the absence of other matter fields in the universe, the scale
factor grows as $a(t)\propto e^{\lambda t}$. Therefore, this
equation reduces to an equation describing propagating waves, whose
solution for wavelengths $\vec{q}^2>9\lambda^2/4$ is
\begin{equation}\label{e:soly}
 y(t,\vec{x})=a(t)^{-\frac{3}{2}}\sum_{\vec{q}}[A\sin(wt+
\vec{q}\cdot\vec{x})+B\cos(wt+\vec{q}\cdot\vec{x})],
\end{equation}
where $w=\pm\sqrt{\vec{q}^2-9\lambda^2/4}$. Hence, the perturbations
$y$ also drops to zero as the universe expands.

From the above solutions of $\Phi$ and $y$, we now can determine the
evolution of the equation of state (\ref{e:eosperfix}). At early
times of the universe when $a$ is very small, the equation of state
$\omega$ is approximately $-2/3$. At late times when $a$ is large,
the perturbations freeze and so the equation of state
$\omega\rightarrow-1$. Hence, due to fluctuations of the brane
universe, the cosmological constant in this brane universe is not
always the same everywhere at early times. It has a wavy profile of
distribution, which leads to $\omega=-2/3$ but not $=-1$. The
equation of state $\omega$ approaches $-1$ only at late times. But,
in both cases, the expansion of the universe accelerates since
$\omega<-1/3$.

\subsection{The origin from tachyon inflation}

The Chaplygin gas model and the tachyon condensate model are
sometimes combined to construct combinatorial model accounting for
inflation \cite{Chimento:2003ta,Herrera:2008us,Herrera:2008bj}. In
this subsection, we show that the fixed BPS D3-brane universe
discussed above can actually find its origin from tachyon inflation
on a non-BPS D4-brane. It can be a solitonic object left behind the
tachyon inflation.

Let us first take a look at the tachyon condensation process on a
non-BPS D$p$-brane in the Minkowski spacetime
\cite{Cline:2003vc,Barnaby:2004dz}. 
Towards the end of tachyon condensation, some solitonic defects form
in regions where $T=0$ and $\dot{T}=0$. It was verified that these
defects correspond to some BPS D$(p-1)$-branes. In other regions,
the tachyon field $T$ gets infinite as $\dot{T}\rightarrow\pm1$
towards the end of condensation.

We now show that the situation is similar in the cosmological
background. As in the usual tachyon inflation model on a non-BPS
D3-brane
\cite{Gibbons:2002md,Fairbairn:2002yp,Choudhury:2002xu,Feinstein:2002aj},
we consider a non-BPS D4-brane coupled to gravity
\begin{eqnarray}\label{e:d4tacinf}
 S_T=\int d^5x\sqrt{-g}V(T)\sqrt{1+g^{\mu\nu}\partial_\mu T
\partial_\nu T},
\end{eqnarray}
where $V(T)$ is the tachyon potential which has the maximum value
$V_m$ at $T=0$ and the minimum value zero as
$T\rightarrow\pm\infty$. There are two usually adopted potentials
derived from string theory: $V=V_m/\cosh(\beta T)$ and
$V=V_m\exp(-\beta^2T^2)$ with $\beta$ a constant.

In the FRW metric, the Friedmann equation is
$H^2=V/(3M_{Pl}^2\sqrt{1-\dot{T}^2})$, with which we derive the
equation of the tachyon field:
\begin{eqnarray}\label{e:d4taceqnt}
 \ddot{T}=-\frac{\sqrt{3V}}{M_{Pl}}\dot{T}(1-\dot{T}^2)^{
\frac{3}{4}}-(\ln V)'(1-\dot{T}^2),
\end{eqnarray}
where the prime denotes the derivative with respect to $T$. For both
the tachyon potentials $V(T)$ mentioned above, $(\ln V)'=0$ at
$T=0$.

Before the inflation starts, the tachyon field fluctuates on the top
of the asymptotic potential $V$. As the fluctuation proceeds, we can
hopefully find some regions where $T=0$, i.e., $(\ln V)'=0$ for the
above two potentials, and $\dot{T}=0$. In terms of Eq.\
(\ref{e:d4taceqnt}), we know that $\ddot{T}=0$ in these regions.
Hence, $T$ and $\dot{T}$ will remain vanishing in these regions,
where the solitonic BPS D3-branes form. In some sense, the tachyon
$T$ is just a tachyonic transverse modulus of the non-BPS D4-brane
and it will evolve into the massless transverse modulus $Y$ of the
BPS D3-branes at the end of tachyon condensation. Thus, these
D3-branes formed at the end of tachyon inflation should be fixed in
the bulk with $\dot{Y}=\dot{T}=0$. They provide the origin of the
fixed brane universe discussed above. By contrast, in other regions
where $T$ and $\dot{T}$ happen to deviate from zero, $\ddot{T}\neq0$
and so $T$, $|\dot{T}|$ will grow with time until
$\dot{T}^2\rightarrow1$, under which $\ddot{T}$ tends to zero and
$|\dot{T}|$ tends to cease growing.

\section{Conclusions}
\label{sec:conclusions}

We have shown that the transverse scalar $Y$ of a BPS D-brane
coupled to gravity is stabilised under the condition of the velocity
$\dot{Y}=0$ in the FRW metric. The vacuum of such a system is in
that the velocity and fluctuations of the D-brane are both
absolutely zero. If the D-brane moves with a non-vanishing velocity
in the bulk, it will slow down to $\dot{Y}=0$. A mechanism
triggering the slow down is the gravitational wave leakage from this
moving D-brane to the bulk. The kinetic energy lost from the slow
down transfers to the internal energy of the D-brane, propelling the
brane universe to expand. The universe undergoes a matter dominated
era and an acceleration era as time increases. This is consistent
with the observation in the original Chplygin gas model. The result
indicates that the BPS D-brane behaves in an opposite way to a
non-BPS D-brane discussed in tachyon inflation model
\cite{Gibbons:2002md,Fairbairn:2002yp,Choudhury:2002xu}: driven by
the asymptotic tachyon potential, the velocity (the time derivative
of the tachyonic scalar) of the non-BPS D-brane in regions away from
defects increases from $0$ to the critical value $1$ as time grows.

On the other hand, if the D-brane has vanishing initial velocity at
a position in the bulk, it will always be kept fixed at this
position. The parameter $6\lambda^2$, which is proportional to the
string coupling $g_s$, provides a cosmological constant. However, in
this universe with only a cosmological constant, the equation of
state evolves from $-2/3$ to $-1$ due to fluctuations of the D-brane
around the fixed location. Interestingly, this kind of fixed brane
universe is a natural product of defects left behind tachyon
inflation on a non-BPS D4-brane.

\section*{Acknowledgements\markboth{Acknowledgements}{Acknowledgements}}

We would like to thank Jianxin Lu and Huanxiong Yang for useful
discussions.

\newpage
\bibliographystyle{JHEP}
\bibliography{b}

\end{document}